\documentclass[11pt, a4paper]{article}
\usepackage[top=1in, bottom=1in, left=1in, right=1in]{geometry} % Sets 1-inch margins
\usepackage{amsmath, amssymb} % Essential for mathematical symbols and equations
\usepackage{graphicx} % For including figures
\usepackage{multirow}
\usepackage{url}
\usepackage{float}
\RequirePackage[backend=biber,style=numeric-comp,sorting=none]{biblatex}
\usepackage{hyperref}
\usepackage{color}
\newcommand{\wwbar}{\mid W\overline{W} \rangle}
\newcommand{\starstate}{\mid \text{Star} \rangle}

\usepackage{orcidlink}

\date{\today}

\begin{document}

\title{Symmetric and asymmetric tripartite states under the lens of entanglement splitting and topological linking}

\author{Sougata Bhattacharyya \orcidlink{0009-0002-9803-4198} \\
%\email[]{a.mandal1.tmsl@ticollege.org}
Department of Astronomy, Astrophysics and Space Engineering, \\Indian Institute of Technology, Indore 453552, India
\and
Sovik Roy \orcidlink{0000-0003-4334-341X} \\
%\email[]{s.roy2.tmsl@ticollege.org}
Department of Mathematics, Techno Main Salt Lake,  \\Techno India Group, EM 4/1, Sector V,  Kolkata  700091, India}

\maketitle
\begin{abstract}

\noindent This work establishes a direct operational connection between the entanglement structures of specific three-qubit states (i.e. multipartite entanglement) and their corresponding topological links. We investigate the symmetric $\wwbar$ state and the asymmetric $\starstate$ state through local projective measurements on individual qubits. The post measurement states are analyzed via their Schmidt rank to characterize residual bipartite entanglement. For the symmetric $\wwbar$ state, measurement of any qubit consistently results in a non-maximally entangled post-measurement state (Schmidt rank 2), analogous to the behavior of a \textit{3-Hopf link} structure, where cutting any ring leaves the remaining two nontrivially linked. On the other hand, the $\starstate$ state exhibits a context-dependent fragility. Its behavior predominantly mirrors that of a \textit{3-link chain}, where severing the central qubit decouples the system, while cutting an outer qubit often preserves a residual link. Crucially, for specific measurement outcomes, the $\starstate$ state also exhibits the defining property of the \textit{Borromean rings}, where the loss of one qubit completely disentangles the remaining two. This analysis provides a concrete interpretation of topological linking structures as a resource for characterizing distributed entanglement and its resilience under local measurement operations, revealing that a single quantum state can contextually embody multiple distinct topological analogues.
\end{abstract}


%\newpage

\section{Introduction}
\label{sec:introduction}

\noindent It is well established that the qubit constitutes the fundamental unit of quantum information. A single qubit can, in principle, exist in a superposition of basis states until a measurement projects it onto a definite outcome. This naturally leads to the question of how much information and correlation structure can be represented in systems comprising multiple qubits \cite{aczel2002}. In quantum information processing, multipartite systems, especially those with qubits distributed among different parties, are of particular importance. The simplest nontrivial case is a bipartite two qubit system, conventionally described as being shared between two parties, Alice and Bob. When such a system is maximally entangled, it gives rise to the well-known Bell states \cite{nielsen2010quantum}, which serve as fundamental resources in a variety of quantum communication and computation protocols \cite{bennett1993teleporting, werner2001all, bennett1992communication, bennett1992quantum, ekert1991}. This scenario can naturally be extended to tripartite systems, where a third party, Charlie, holds an additional qubit. In such systems, the structure of entanglement becomes significantly richer \cite{horodecki2009}. In particular, two inequivalent classes of genuinely tripartite entangled states have received considerable attention in this regard, viz. the Greenberger–Horne–Zeilinger ($GHZ$) state and the $W$ state. As shown by Dür et al. \cite{dur2000}, these two classes of states are inequivalent under stochastic local operations and classical communication (SLOCC), meaning no state belonging to one class can be converted to a state that belongs to the other class via SLOCC. This distinction highlights the complexity of entanglement in multipartite settings.\\

\noindent In parallel, another rich and independently developed concept in modern physics is topological entanglement, especially as studied in knot theory, which classifies the embedding of loops in three-dimensional space \cite{adams1994}. Traditionally, quantum entanglement and topological entanglement have been studied in separate domains, the former as a resource in quantum information science \cite{nielsen2010quantum,cleve1999,raussendorf2001}, and the latter in mathematical topology \cite{adams1994,rolfsen1976}. However, a growing body of work explores potential correspondences between these two forms of entanglement. A central question is whether the intricate correlation structures of multipartite quantum states can faithfully be captured by topological links, such that operations on quantum states (e.g., measurements) have well-defined analogues in topological manipulations (such as cutting a particular component of a link).\\\\

\noindent P. K. Aravind pioneered this line of investigation, demonstrating suggestive connections between entangled quantum states and topological link structures  \cite{aravind1997}.  Aravind argued for a correspondence between the three qubit $GHZ$ state and the Borromean rings, a classical example of a topological link where the removal of any one component unlinks the remaining two \cite{adams1994,aravind1997}. In this work, we focus on two distinct three qubit pure states that exhibit contrasting entanglement properties viz., (a) The $\wwbar$ state, which is the equal superposition of the standard $\mid W \rangle$ state and its spin-flipped counterpart $\mid \overline{W} \rangle$, and (b) the $\starstate$ state, which is a special type of graph state \cite{cao2020fragility, roy2023exploring, bhattacharjee2025teleportation}. The $\wwbar$ state is symmetric under particle exchange and exhibits quantum coherence and correlations at the single-qubit, bipartite, and tripartite levels. In contrast, the $\starstate$ state, although also displaying coherence and correlations across all levels, is inherently asymmetric in its distribution of quantum correlations. From the perspective of entanglement splitting and topological link analogues, these two states provide fertile ground for analysis. In particular, their algebraic and correlation structures may correspond to distinct types of topological links, potentially offering new insight into the classification and manipulation of multipartite entangled states.\\\\

\noindent Our primary motivation arises from the increasing need, in the so-called \textit{Quantum 2.0 era}, to deepen our understanding of decoherence in complex quantum systems and explore strategies for its mitigation. Adopting a purely mathematical perspective, we examine how multipartite entanglement structures can be interpreted as well as potentially stabilized through topological insights. This cross-disciplinary framework, which bridges quantum information theory and knot theory, offers promising avenues for the analysis and design of robust quantum systems. The paper is structured as follows: section 2 highlights the theoretical preliminaries, followed by our main results which involve $\wwbar$ and $\starstate$ in Section 3. In Section 4, we analyze these two states in detail. Section 5 discusses the broader implications of our findings, with concluding remarks and future research directions presented in Section 6.

\section{Theoretical Preliminaries}
\label{sec:preliminaries}

In this section, we outline the fundamental concepts necessary for our analysis of topological links in relation to entanglement splitting within specific tripartite states. We first review the quantum information framework, focusing on multipartite entanglement, the formalism of projective measurements, and the use of Schmidt decomposition as a means of quantifying bipartite entanglement. We then present the key topological notions of knots and links, which serve as the analogues underpinning our interpretation of the quantum insights.

\subsection{Review of tripartite Entanglement}

\noindent A quantum state composed of two or more subsystems is said to be \textit{entangled} if it cannot be expressed as a product state of its constituents. In multipartite scenario such as those involving three parties, the structure of entanglement becomes considerably richer and more intricate than in the bipartite case. Genuine tripartite entanglement manifests in distinct forms, known as \textit{entanglement classes}, which cannot be transformed into one another via Stochastic Local Operations and Classical Communication (SLOCC) \cite{dur2000,bennett1996}.
\\\\
\noindent Among three qubit systems, the most prominent representatives are the $GHZ$ class and the $W$ class. The canonical $GHZ$ state,
\begin{eqnarray}
\label{ghz}
\mid \text{GHZ} \rangle = \frac{1}{\sqrt{2}}(\mid 000 \rangle + \mid 111 \rangle),
\end{eqnarray}
is marked by its high fragility i.e. a measurement on any single qubit in the computational basis collapses the entire system into a separable product state \cite{greenberger1990}. In contrast, the canonical $W$ state,
\begin{eqnarray}
    \mid \text{W} \rangle = \frac{1}{\sqrt{3}}(\mid 001 \rangle + \mid 010 \rangle + \mid 010 \rangle)
\end{eqnarray}
is more robust which means any projective measurement on a single qubit leaves the remaining pair in a nonmaximally entangled state \cite{dur2000}. Both of these states have been experimentally realized, and their usefulness, whether in prototypical or non-protypical forms, for quantum information processing is now firmly established \cite{bouwmeester1999observation, pan2000experimental, eibl2004experimental, agrawal2006perfect}. In this work, we focus on a distinct class of tripartite states that represent another frontier in the study of multipartite quantum systems and hold notable importance for quantum information processing. Specifically, we consider the $\wwbar$ and $\starstate$ states, whose relevance has already been demonstrated in earlier studies \cite{cao2020fragility, bhattacharjee2025teleportation, roy2025environment, radhakrishnan2024entanglement, roy2025dephasing}.
\subsection{Projective Measurements}

\noindent A projective measurement is described by an \textit{observable}, a Hermitian operator $M$ on the state space of the system. The observable has a spectral decomposition $M = \sum_m m P_m$, where $m$ are the possible measurement outcomes and $P_m$ is the projector onto the eigenspace corresponding to $m$. The projectors are Hermitian, idempotent ($P_m^2 = P_m$), and complete ($\sum_m P_m = I$) \cite{nielsen2010quantum}.\\

\noindent Consider a three-qubit state $\mid \psi_{ABC} \rangle$. A projective measurement on qubit $A$ in its computational basis $\{ \mid 0 \rangle, \mid 1 \rangle \}$ is described by the projectors $P^A_0 = \mid 0 \rangle\langle 0 \mid_A \otimes I_B \otimes I_C$ and $P^A_1 = \mid 1 \rangle\langle 1 \mid_A \otimes I_B \otimes I_C$. The probability of obtaining outcome $k$ is given by the Born rule \cite{nielsen2010quantum}:
\begin{equation}
p(k) = \langle \psi_{ABC} \mid P^A_k \mid \psi_{ABC} \rangle.
\end{equation}
The state of the system \textit{after} the measurement, conditional on outcome $k$, is the \textit{post-measurement state}:
\begin{equation}
\mid \psi^{(k)}_{BC} \rangle = \frac{P^A_k \mid \psi_{ABC} \rangle}{\sqrt{p(k)}}.
\end{equation}
This resulting state is a bipartite state on the unmeasured qubits $B$ and $C$, whose entanglement properties we wish to analyze.

\subsection{Schmidt Rank}

\noindent The Schmidt decomposition is a powerful tool for analyzing bipartite pure states. Any pure state $\mid \psi \rangle_{BC}$ of two systems $B$ and $C$ can be written in the form \cite{peres1995}:
\begin{equation}
\mid \psi \rangle_{BC} = \sum_{i=1}^R \lambda_i \mid u_i \rangle_B \otimes \mid v_i \rangle_C,
\end{equation}
where $\{ \mid u_i \rangle_B \}$ and $\{ \mid v_i \rangle_C \}$ are orthonormal bases for systems $B$ and $C$ respectively, $\lambda_i$ are non-negative real Schmidt coefficients satisfying $\sum_i \lambda_i^2 = 1$, and the number $R$ of non-zero $\lambda_i$ is called the \textbf{Schmidt rank}.\\

\noindent The Schmidt rank provides a coarse-grained measure of entanglement:
\begin{itemize}
    \item A Schmidt rank of 1 ($R=1$) indicates the state is \textit{separable} (product state).
    \item A Schmidt rank greater than 1 ($R>1$) indicates the state is \textit{entangled}.
    \item A Schmidt rank of 2 ($R=2$) with $\lambda_1 = \lambda_2 $ indicates a \textit{maximally entangled} bipartite state (e.g., a Bell state) \cite{einstein1935}.
\end{itemize}
In our analysis, we will calculate the Schmidt rank of the post-measurement state $\mid \psi^{(k)}_{BC} \rangle$ to determine if the unmeasured qubits $B$ and $C$ remain entangled after the measurement on $A$.

\subsection{Fundamentals of Knots and Links}

\noindent Knot theory is the mathematical study of closed curves in three-dimensional space. A \textbf{knot} is a single closed curve that is embedded in 3D space and cannot be untangled to form a simple loop (the \textit{unknot}). A \textbf{link} is a collection of one or more knots that may be intertwined with each other \cite{adams1994}.\\

\noindent Two links are considered equivalent if one can be smoothly deformed into the other without cutting or passing one strand through another. Links are often characterized by their \textit{invariants}, properties that remain unchanged under such smooth deformations. Simple invariants include the number of components and how they are linked, quantified for two-component links (called \textit{2-links}) \cite{aravind1997} by the \textit{linking number} \cite{rolfsen1976}. The key topological operation we will use as an analogue to quantum measurement is the \textit{cutting} or \textit{deletion} of a link component. The central question is: if one component of a multi-component link is removed, what is the nature of the resulting link between the remaining components?\\

\begin{itemize}
    \item The \textit{Hopf link} is the simplest non-trivial 2-link. If one of its two rings is cut and removed, the remaining ring becomes an unknot, i.e., it is completely \textit{unlinked} \cite{adams1994}.

    \begin{figure}[H]
    \centering
    \includegraphics[scale=0.15]{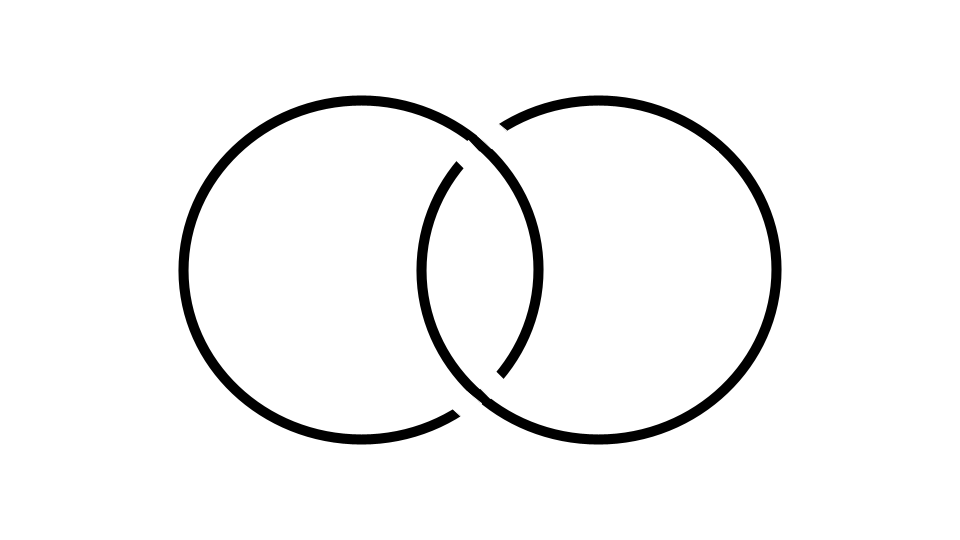}
    \caption{Hopf Link}
    \end{figure}
    
    \item The \textit{3 Hopf link} is a non-trivial 3-link. Its defining property is that all three rings are \textit{pairwise} directly linked. \textit{Cutting and removing any one ring} results in a Hopf link (a non-trivial 2-link) between the remaining two rings \cite{adams1994}.

    \begin{figure}[H]
    \centering
    \includegraphics[scale=0.15]{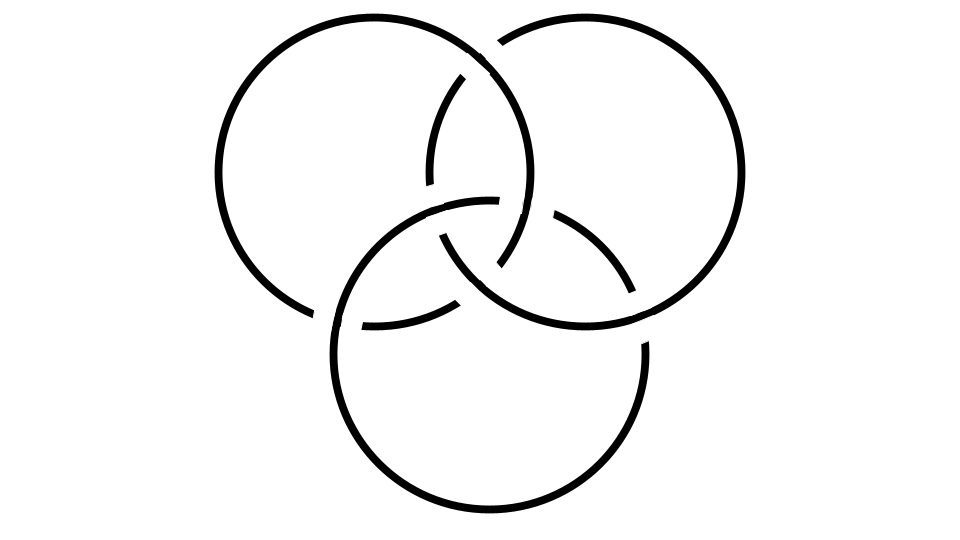}
    \caption{3 Hopf Rings}
    \end{figure}
    
    \item The \textit{Borromean rings} are a famous 3-link. Their defining property is that no two rings are directly linked; each pair is separable. However, all three are topologically bound. \textit{Cutting and removing any one ring} causes the other two to fall apart, becoming two separate unknots (a \textit{trivial 2-link}) \cite{aravind1997}.
    
    \begin{figure}[H]
    \centering
    \includegraphics[scale=0.13]{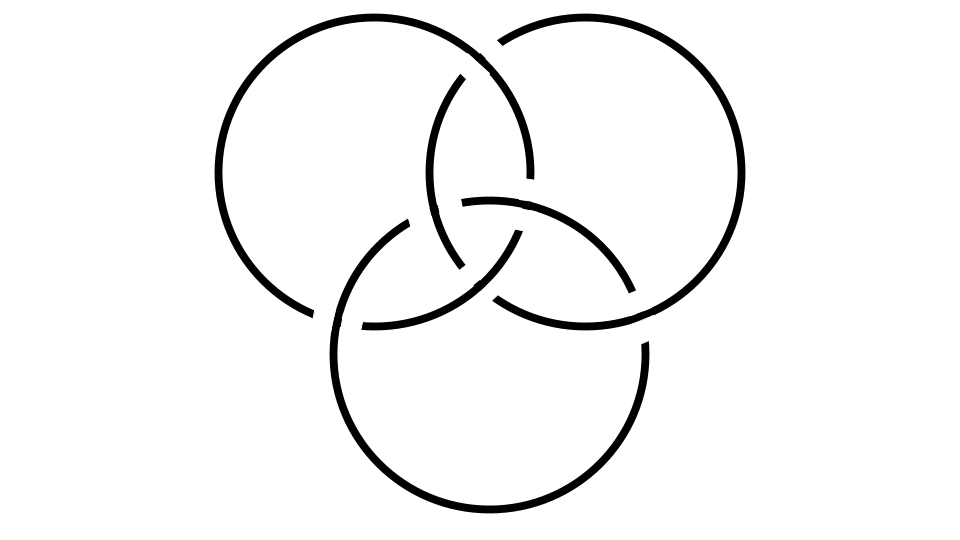}
    \caption{Borromean Rings}
    \end{figure}
    
    \item The \textit{3 Link Chain} consists of three components. Its key property is its \textit{asymmetry}. Cutting a specific component (e.g., the central one in a certain presentation) may completely unlink the system. In contrast, cutting a different component might leave behind a non-trivial link (e.g., a Hopf link) between the remaining two \cite{adams1994}.

    \begin{figure}[H]
    \centering
    \includegraphics[scale=0.15]{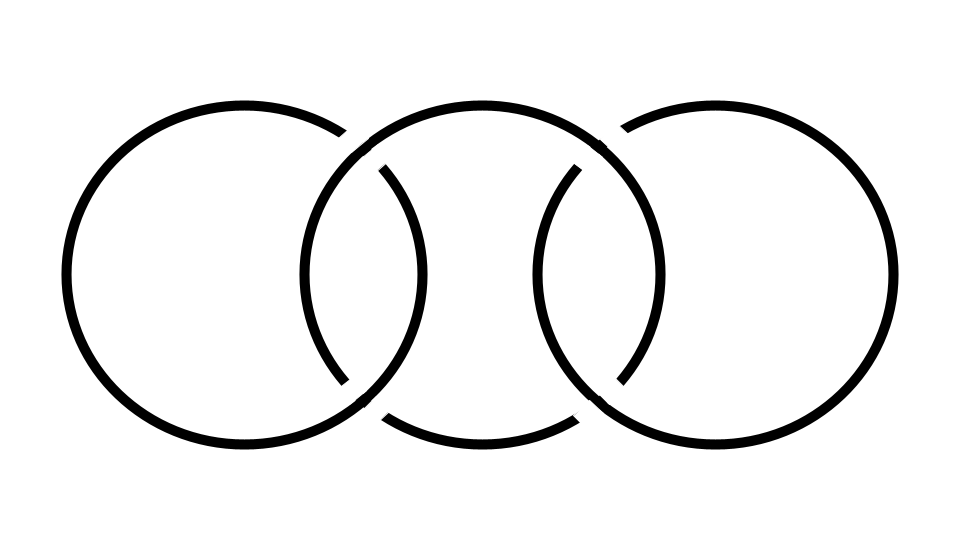}
    \caption{3 Link Chain}
    \end{figure}
    
\end{itemize}

\section{Overview of Main Results}
We now turn our attention to two particular three-qubit states, the $\wwbar$ and $\starstate$ states, which display rich yet contrasting patterns of entanglement and form the central focus of this study. Their distinctive algebraic features render them especially suitable for our investigation, wherein we explore their entanglement structure through the operational framework of \textit{local projective measurement}.\\\\

\noindent The $\wwbar$ state is defined as 

\begin{eqnarray}
\wwbar = \frac{1}{\sqrt{2}} (\mid W \rangle + \mid \overline{W} \rangle),
\end{eqnarray}

where $\mid \overline{W} \rangle $ is the spin-flipped version of $W$ state and is given as

\begin{eqnarray}
\mid W \rangle = \frac{1}{\sqrt{3}}(\mid 001 \rangle + \mid 010 \rangle + \mid 100 \rangle)
\end{eqnarray}

and 
\begin{eqnarray}
\mid \overline{W} \rangle = \frac{1}{\sqrt{3}}(\mid 011 \rangle + \mid 101 \rangle + \mid 110 \rangle)
\end{eqnarray}

The $\starstate$ is given by 

\begin{eqnarray}
\starstate = \frac{1}{2}(\mid 000 \rangle + \mid 100 \rangle + \mid 101 \rangle + \mid 111 \rangle )
\end{eqnarray}

\noindent Although our main focus in this article is to study the topological structures of two tripartite states, we briefly discuss here the experimental realization of the above states. The experimental generations of the $\wwbar$ and $\starstate$ states have been demonstrated using all optical setups by Cao et al. \cite{cao2020fragility}.\\\\

\noindent The $\wwbar$ state can be obtained from a higher-order Spontaneous Parametric Down Conversion (SPDC) process, where two photon pairs are simultaneously created in a beamlike type-II $\beta$ Barium Borate (BBO) crystal. The four photons are collected into a single spatial mode and overlapped on a Polarizing Beam Splitter (PBS), followed by spectral filtering (3-nm bandwidth) and splitting using three non-polarizing beam splitters. Post selection of fourfold coincidences yields the four photon Dicke state $\mid {D_4^{(2)}\rangle}$, from which a projection of one qubit onto $\frac{1}{\sqrt{2}}(\mid 0\rangle + \mid 1\rangle)$ generates the desired $\wwbar$ state. The $\starstate$ state, on the other hand, is constructed by preparing two nonmaximally entangled EPR (Einstein-Podolsky-Rosen) pairs using a \textit{sandwiched} beamlike type-II BBO source (BBO-HWP-BBO) \footnote{HWP stands for \textit{Half Wave Plate}.}, with additional spatial and temporal compensation. After suitable local unitaries, the pairs are tuned into the form  $\frac{1}{\sqrt{3}}(\mid 00\rangle+\mid 10\rangle+\mid 11\rangle)$, corresponding to the ratio $\frac{\cos^2{\theta}}{\sin^2{\theta}} = 6.8554$. The two pairs are interfered with on a PBS acting as a parity-check gate, with indistinguishability enhanced using a $2$-nm filter to improve Hong-Ou-Mandel visibility. A subsequent projection of one photon onto $\frac{1}{\sqrt{2}}(\mid 0 \rangle + \mid 1\rangle)$, together with a swap of qubits $3$ and $4$, results in the preparation of the three-qubit $\starstate$ state \cite{cao2020fragility}. 
\\\\
\subsection{Topological Analysis:}
\noindent For each state, we projectively measure one qubit in the computational basis and analyze the resulting two-qubit post-measurement state. The key quantities of interest are the Schmidt rank and the degree of entanglement of this residual state, which diagnose the presence and strength of bipartite entanglement between the unmeasured qubits. Our central finding is that the pattern of entanglement degradation revealed by these measurements maps onto the splitting behaviour of topological links.\\\\
\begin{itemize}
    \item For the $\wwbar$ state, the measurement of any qubit \textit{consistently} results in a non-maximally entangled post-measurement state (\textit{Schmidt rank} 2). This is analogous to the cutting of one component of a structure of three \textit{Hopf links} \cite{aravind1997};  the remaining two components are still linked, but not as \textit{tightly} as in their original configuration. The state exhibits robust, persistent pairwise entanglement.

    \item For the $\starstate$ state, the result is more nuanced and reveals a hybrid topological character. The state's behavior under measurement is predominantly analogous to that of a \textit{3 link chain} \cite{adams1994}.\\

\noindent (1) Measurement on the central qubit severs all entanglement (\textit{Schmidt rank }1), analogous to cutting the central ring of the 3 link chain, which unlinks the two outer rings.\\\\
\noindent (2)  Measurement on an outer qubit most often leaves the other outer qubit and the central qubit entangled (\textit{Schmidt rank} 2), analogous to cutting an outer ring and leaving a Hopf link between the central ring and the remaining outer ring.\\
\end{itemize}
\noindent Crucially, for specific measurement outcomes ($\mid 0 \rangle$ on qubit $A$ or $\mid 1 \rangle$ on qubit $B$), the $\starstate$ state exhibits the defining property of the \textit{Borromean rings} \cite{aravind1997}. In these cases, the measurement completely destroys all entanglement (\textit{Schmidt rank} 1), revealing that the lost qubit was essential for mediating the tripartite correlation precisely as cutting any ring of the\textit{ Borromean rings} completely unlinks the remaining rings.\\\\
\noindent This analysis provides a clear, operational bridge between quantum information theory and topology. It suggests that the resilience and \textit{configuration} of distributed entanglement after local measurements is encoded in the topology of an analogous link. The $\starstate$ state, in particular, offers a fascinating example where a single quantum state can contextually embody the properties of two distinct topological structures i.e. (a) the 3 link chain and (b) the Borromean rings, depending on the measurement context. 

\section{Analysis of symmetric and asymmetric tripartite states:}
\subsection{Analysis of the $\wwbar$ State}
\label{sec:wwbar}

\noindent This section presents a summary of the entanglement properties of the $\wwbar$ state, as revealed through local projective measurements. A comprehensive mathematical derivation of all results can be found in Appendix \ref{app:wwbar_calc}.

\subsubsection{Summary of Measurement Outcomes}

\noindent The $\wwbar$ state exhibits a high degree of permutation symmetry \cite{agrawal2002}. Consequently, the outcome of a projective measurement on any single qubit is identical, regardless of whether qubit $A$, $B$, or $C$ is chosen. The measurement is performed in the computational basis $\{ \mid 0 \rangle, \mid 1 \rangle \}$. The results, detailed in Table \ref{tab:wwbar_results}, show that for any measurement, one outcome yields a post-measurement state with \textit{Schmidt rank} 2, indicating residual bipartite entanglement between the unmeasured qubits. The other outcome also yields a state with\textit{ Schmidt rank} 2, confirming a consistent pattern of robust entanglement.

\begin{table}[h!]
\centering
\label{tab:wwbar_results}
\begin{tabular}{|c|c|c|c|c|}
\hline
\textbf{Qubit Measured} & \textbf{Outcome} & \textbf{Probability} & \textbf{Post-Measurement State ($\mid \psi \rangle$)} & \textbf{Schmidt Rank} \\ \hline
\multirow{2}{*}{Any ($A$, $B$, or $C$)} & $\mid 0 \rangle$ & $\frac{1}{2}$ & $\frac{1}{\sqrt{3}}\left( \mid 01 \rangle + \mid 10 \rangle + \mid 11 \rangle \right)$ & 2 \\ \cline{2-5} 
 & $\mid 1 \rangle$ & $\frac{1}{2}$ & $\frac{1}{\sqrt{3}}\left( \mid 00 \rangle + \mid 01 \rangle + \mid 10 \rangle \right)$ & 2 \\ \hline
\end{tabular}
\caption{Summary of projective measurement results on the $\wwbar$ state. Due to permutation symmetry, the result is identical for a measurement on qubit $A$, $B$, or $C$.}
\end{table}

\subsubsection{Interpretation of Results}

\noindent The key finding from this analysis is the \textit{robustness} of entanglement in the $\wwbar$ state. Unlike the GHZ state, which becomes completely separable upon any single-qubit measurement, the $\wwbar$ state retains bipartite entanglement for all possible measurement outcomes \cite{dur2000}. The \textit{Schmidt rank} is consistently 2, signifying that the unmeasured qubits are never found in a separable product state. However, the entanglement is nonmaximal. The two non-zero eigenvalues of the reduced density matrix for the post-measurement states are $\lambda = \frac{3 \pm \sqrt{5}}{6}$ (see Appendix \ref{app:wwbar_calc}), which are not equal. This distinguishes the residual entanglement from that of a maximally entangled Bell state, where the eigenvalues would be equal ($\lambda_1 = \lambda_2 = \frac{1}{2}$).\\

\subsubsection{Topological Analogue: Three Hopf Links}

\noindent The entanglement properties of the $\wwbar$ state under measurement, finds a natural topological analogue in a structure of  \textit{3 Hopf rings} \cite{aravind1997, kauffman2016}. The Hopf link is the simplest non-trivial link, consisting of two interlocked rings. As already been discussed in the earlier section, its defining topological property is that if one ring is \textit{cut and removed}, the other ring is released as a trivial unknot; the linking is completely destroyed \cite{adams1994}.\\

\noindent Now, consider a symmetric configuration of three rings, labeled A, B, and C, where each \textit{pair} (A-B, A-C, B-C) is linked in the manner of a Hopf link. This is not the Borromean ring (where no pair is linked), but a different structure where each pair is individually linked. The analogue to our quantum measurement is the \textit{cutting and removal} of one ring \cite{aravind1997}. If we cut and remove ring A,

\begin{itemize}
    \item The direct Hopf link between rings B and C \textit{remains intact}. They are still linked.
    \item The system is reduced from a three-component link to a two-component link (B-C).
\end{itemize}

\noindent This mirrors our quantum result:
\begin{itemize}
    \item \textbf{Measurement ($=$ Cutting):} Projectively measuring (and thereby disconnecting) any qubit.
    \item \textbf{Residual Entanglement ($=$ Residual Linking):} The remaining two qubits are always entangled (\textit{Schmidt rank} 2).
     This is the quantum counterpart of the Hopf link between the two remaining rings persisting after the third is removed. The nonmaximal nature of the entanglement may correspond to a specific type of nontrivial, but not maximal, linking between the remaining rings.
\end{itemize}
Similarly, if one cuts the ring B the reduced system is two component link (A-C) and if the ring C is cut the reduced system is two component link (A-B).
\noindent The consistency of this result, regardless of which qubit is measured (due to the state's symmetry), strengthens the analogy to the symmetric three Hopf-link structure. The $\wwbar$ state exhibits a form of \textit{robust pairwise entanglement} that is preserved under measurement of a third party, much like the pairwise linking in the three Hopf-link configuration is preserved when one component is removed \cite{aravind1997}.

\begin{figure}[H]
\centering
\includegraphics[scale=0.15]{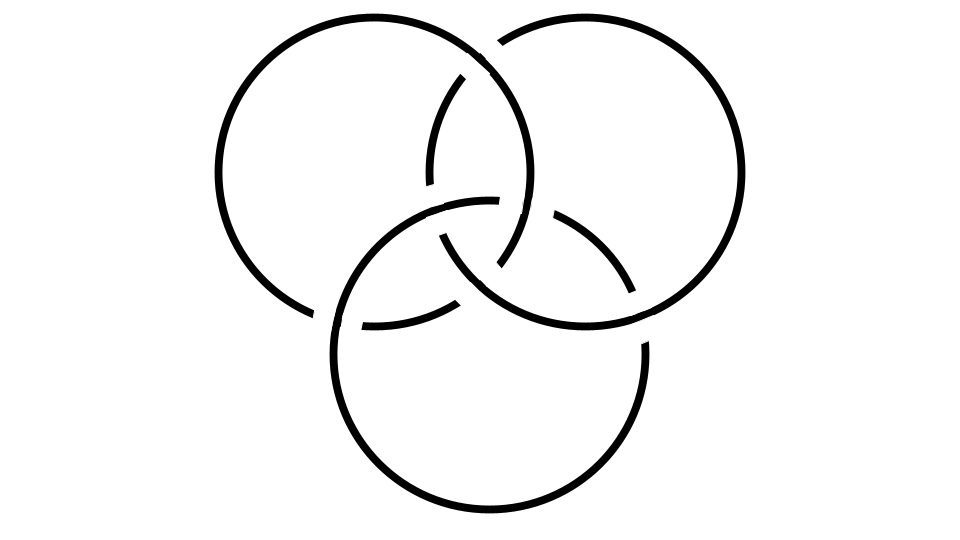}
\caption{3 Hopf Rings}
\end{figure}

\subsection{Analysis of the $\starstate$ State}
\label{sec:star}

\noindent This section presents an analysis of the entanglement properties of the $\starstate$ state, as revealed through local projective measurements. In contrast to the symmetric $\wwbar$ state, the $\starstate$ state exhibits a distinct asymmetry between its \textit{central qubit} (designated as qubit $C$) and its \textit{outer qubits} (or \textit{peripheral qubits}) ($A$ and $B$). A comprehensive mathematical derivation of all results can be found in Appendix \ref{app:star_calc}.

\subsubsection{Summary of Measurement Outcomes}

The results of projective measurements on the $\starstate$ state are summarized in Table \ref{tab:star_results}. The behavior is highly dependent on which qubit is measured.

\begin{table}[h!]
\centering
\label{tab:star_results}
\begin{tabular}{|c|c|c|c|c|}
\hline
\textbf{Qubit Measured} & \textbf{Outcome} & \textbf{Probability} & \textbf{Post-Measurement State ($\mid \psi \rangle$)} & \textbf{Schmidt Rank} \\ \hline
\multirow{2}{*}{Central ($C$)} & $\mid 0 \rangle_C$ & $\frac{1}{2}$ & $\frac{1}{\sqrt{2}} \left( \mid 00 \rangle + \mid 10 \rangle \right)_{AB}$ & 1 \\ \cline{2-5} 
 & $\mid 1 \rangle_C$ & $\frac{1}{2}$ & $\frac{1}{\sqrt{2}} \left( \mid 10 \rangle + \mid 11 \rangle \right)_{AB}$ & 1 \\ \hline
\multirow{2}{*}{Outer ($A$)} & $\mid 0 \rangle_A$ & $\frac{1}{4}$ & $\mid 00 \rangle_{BC}$ & 1 \\ \cline{2-5} 
 & $\mid 1 \rangle_A$ & $\frac{3}{4}$ & $\frac{1}{\sqrt{3}} \left( \mid 00 \rangle + \mid 01 \rangle + \mid 11 \rangle \right)_{BC}$ & 2 \\ \hline
\multirow{2}{*}{Outer ($B$)} & $\mid 0 \rangle_B$ & $\frac{3}{4}$ & $\frac{1}{\sqrt{3}} \left( \mid 00 \rangle + \mid 10 \rangle + \mid 11 \rangle \right)_{AC}$ & 2 \\ \cline{2-5} 
 & $\mid 1 \rangle_B$ & $\frac{1}{4}$ & $\mid 11 \rangle_{AC}$ & 1 \\ \hline
\end{tabular}
\caption{Summary of projective measurement results on the $\starstate$ state. The central qubit is $C$; qubits $A$ and $B$ are outer or peripheral qubits.}
\end{table}

\subsubsection{Interpretation of Results}

The key finding from this analysis is the \textit{asymmetric} and \textit{context-dependent} nature of entanglement in the $\starstate$ state.

\begin{itemize}
    \item \textbf{Measurement on the Central Qubit:} Projective measurement on the central qubit $C$ \textit{always} severs the entanglement between the outer qubits. Regardless of the outcome ($\mid 0 \rangle_C$ or $\mid 1 \rangle_C$ (with probability $\frac{1}{2}$), the resulting post-measurement state is a separable product state (\textit{Schmidt rank} 1). This signifies that the central qubit is crucial for mediating the entanglement between the outer qubits such that its removal completely disconnects them.

    \item \textbf{Measurement on an Outer Qubit:} The outcome of measuring an outer qubit is probabilistic and exhibits a further asymmetry between qubits $A$ and $B$. For qubit $A$, the outcome $\mid 0 \rangle_A$ (probability $\frac{1}{4}$) yields a separable state for $BC$, while the outcome $\mid 1 \rangle_A$ (probability $\frac{3}{4}$) leaves $BC$ entangled. For qubit $B$, the outcome $\mid 0 \rangle_B$ (probability $\frac{3}{4}$) yields an entangled state for $AC$, while the outcome $\mid 1 \rangle_B$ (probability $\frac{1}{4}$) leaves $AC$ separable. This demonstrates that the entanglement is not uniformly distributed; the persistence of correlation between the remaining qubits depends on both \textit{which} outer qubit is measured and the specific measurement outcome.
\end{itemize}

\subsubsection{Topological Analogue: The 3 Link Chain with Borromean Character}

The entanglement properties of the $\starstate$ state under measurement are most precisely captured by the \textit{3 link chain}, but its behavior reveals a subtle and profound nuance, for specific measurement outcomes, it exhibits the defining property of the \textit{Borromean rings} \cite{adams1994, aravind1997}.\\

The analogue to our quantum measurement is the \textit{cutting and removal} of one component:
\begin{itemize}
    \item \textbf{The 3 Link Chain Behavior:} The predominant behavior mirrors that of a 3 link chain. In this structure, the central ring is topologically crucial.
    \begin{itemize}
        \item \textbf{Cutting the Central Component ($=$ Measuring Qubit C):} Removing the central ring causes the two outer rings to become completely unlinked. This corresponds to measuring qubit $C$, which always severs the entanglement between $A$ and $B$ (\textit{Schmidt rank} 1).
        \item \textbf{Cutting an Outer Component ($=$ Most Outcomes):} Removing one outer ring from a 3 link chain leaves the other outer ring and the central ring non-trivially linked (in a Hopf link). This corresponds to the most probable outcomes of measuring an outer qubit, $\mid 1 \rangle_A$ and $\mid 0 \rangle_B$, which leave the central qubit entangled with the other outer qubit (\textit{Schmidt rank} 2).
    \end{itemize}
    
    \begin{figure}[H]
    \centering
    \includegraphics[scale=0.15]{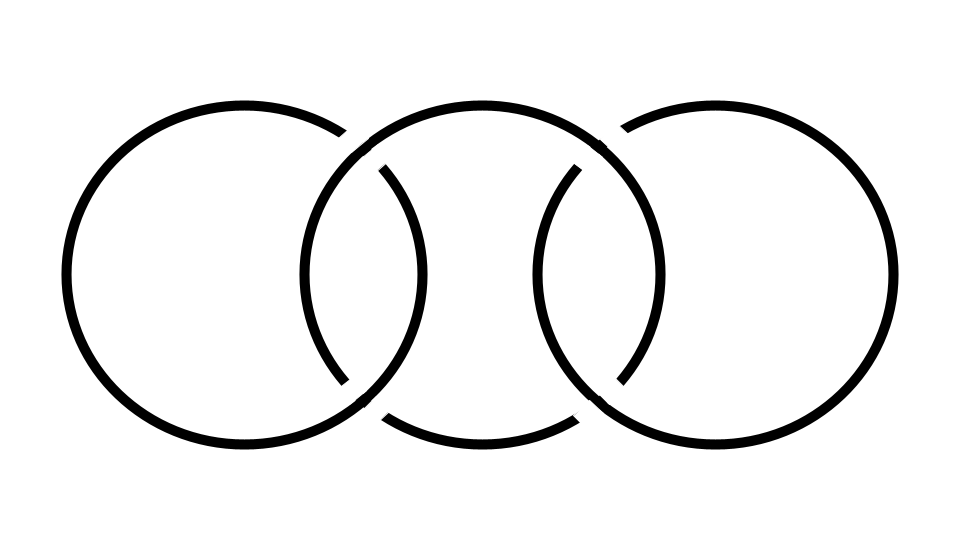}
    \caption{3 Link Chain}
    \end{figure}

    \item \textbf{The Borromean Rings Behavior:} Crucially, for the specific outcomes $\mid 0 \rangle_A$ and $\mid 1 \rangle_B$, the state behaves like the \textit{Borromean rings}. The Borromean rings have the defining property that \textit{no two rings are individually linked}; their entanglement is purely tripartite. Therefore, cutting and removing any one ring causes the other two to fall apart, becoming completely unlinked. This is exactly what happens upon measuring qubit $A$ and getting $\mid 0 \rangle_A$ or measuring qubit $B$ and getting $\mid 1 \rangle_B$. In these cases, the entire multipartite entanglement is destroyed, and the remaining two qubits are found in a separable product state (Schmidt rank 1). The loss of one specific qubit (with a specific outcome) severs all correlations, revealing that those particular qubits were not directly linked in a pairwise fashion.
  
\end{itemize}

\noindent This analysis provides a sophisticated topological interpretation: the $\starstate$ state \textit{is not a single, static link but a structure whose splitting behavior under measurement oscillates between that of a 3 link chain and the Borromean rings}. This depends critically on which qubit is measured and the specific outcome obtained. It demonstrates that the Borromean property—the hallmark of genuine tripartite entanglement with no pairwise links—can exist contextually within a state that also exhibits robust pairwise (Hopf-like) connections.

\begin{figure}[H]
\centering
\includegraphics[scale=0.15]{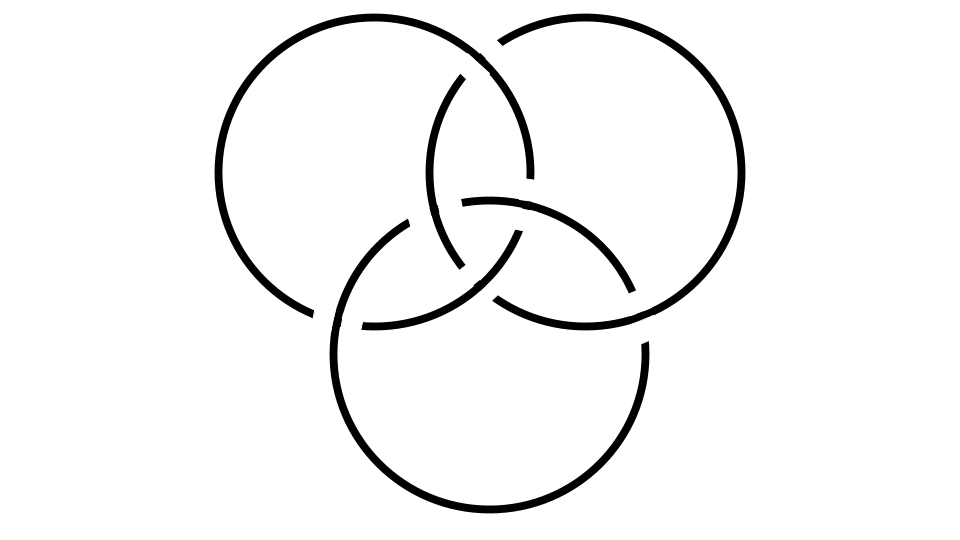}
\caption{Borromean Rings}
\end{figure}

\section{Discussion and Implications}

\noindent The operational analysis performed in this paper, linking local projective measurements to the cutting of topological links, provides a concrete and intuitive framework for understanding complex multipartite entanglement. The consistent correspondence we have uncovered between the \textit{Schmidt rank} of post-measurement states and the \textit{linking of residual topological components} is not merely a formal curiosity; it offers profound insights into the nature of distributed quantum correlations and suggests practical implications for quantum information science.

\subsection{Operationalizing Topological Entanglement}

\noindent Traditionally, the connection between quantum and topological entanglement has often been abstract, relying on formal mathematical similarities or the use of topological invariants like knot polynomials to classify states \cite{kauffman2016, li2008}. Our approach is fundamentally different and more operational. We use a standard quantum information protocol i.e. local measurement, as a direct analogue for a topological operation cutting a component. The resulting pattern of entanglement degradation (quantified by the \textit{Schmidt rank}) then maps directly onto the splitting behavior of an analogous link. This provides a \textit{procedural interpretation} of topological linking: a link diagram can be viewed as a blueprint predicting the resilience of quantum correlations against the loss of a subsystem. The Hopf link, for instance, represents robust but nonmaximal pairwise entanglement that survives the measurement of a third party, precisely as seen in the $\wwbar$ state.

\subsection*{The $\starstate$ State: A Hybrid Topological Entity}

\noindent The $\starstate$ state presents a fascinating case study. Our results demonstrate that its entanglement structure cannot be uniquely assigned to a single, static topological link. Instead, its behavior is \textit{contextual}, depending on which qubit is measured and the specific outcome obtained. Predominantly, \textit{it behaves like a 3 link chain}, where the central qubit is essential for mediating entanglement between the outer qubits. However, for specific measurement outcomes ($|0\rangle_A$ and $|1\rangle_B$), it exhibits the quintessential property of the Borromean rings i.e. the complete severance of all residual entanglement upon the loss of one component \cite{aravind1997}.\\

\noindent This contextual behavior suggests that the $\starstate$ state is a quantum embodiment of a \textit{hybrid topological entity}. It possesses the generic splitting properties of a 3 link chain but retains the potential for a \textit{Borromean catastrophe} (complete disentanglement) under specific conditions. This nuance is in tune with the standard SLOCC classification \cite{dur2000} by showing that states within a single class can exhibit operationally distinct, measurement-contextual entanglement features that are topologically interpretable.

\subsection{Implications for Quantum Information Processing}

\noindent This topological analogy is not just a descriptive tool rather it has potential implications for designing and understanding quantum protocols.

\begin{itemize}
    \item \textbf{Resource State Selection:} The analysis provides a clear criterion for selecting resource states for tasks in distributed quantum computation and networking. If a protocol requires that entanglement between any two nodes persists even if a third node fails or is measured (e.g., for robust quantum secret sharing \cite{cleve1999}), a state with a \textit{$3$ Hopf links} structure like $\wwbar$ is ideal. Conversely, if the protocol requires that the compromise of a central party (e.g., a server) completely severs connections between clients, a state with a central link component like $\starstate$ is more suitable.
    \item \textbf{Measurement-Based Quantum Computing (MBQC):} In MBQC, the initial highly-entangled resource state is processed by adaptive local measurements \cite{raussendorf2001}. The resilience of entanglement to these measurements directly impacts the computation's fault tolerance and complexity. Our topological analogy offers an intuitive way to categorize resource states based on how their entanglement graph is pruned by measurements, potentially aiding in the search for new resource states with desirable robustness properties.
\end{itemize}

\subsection{Limitations and Scope of the Analogy}

Though powerful, this operational analogy has its own limits. It is a coarse-grained tool (i.e. a theoretical or computational method used to simplify a complex, many-body quantum system by focusing only on its most important, large-scale (or \textit{coarse}) properties), primarily sensitive to the Schmidt rank (entanglement vs. separability) rather than the finer details of entanglement measures like concurrence or entanglement entropy \cite{wootters1998,plenio2007}. For instance, the analogy distinguishes a maximally entangled Bell state (\textit{Schmidt rank 2}, $\lambda_1=\lambda_2$) from the non-maximally entangled states resulting from measuring $\wwbar$ (\textit{Schmidt rank 2}, $\lambda_1 \neq \lambda_2$), but the topological interpretation of this difference in strength of linking is less immediately clear. Furthermore, the analogy is currently applied to specific states and measurements; a comprehensive dictionary mapping all local operations to topological manipulations remains an open and exciting challenge for future work.

\section{Conclusion}

\subsection{Summary of Findings}

\noindent In this work, we have established a direct operational bridge between the entanglement structure and topological links of some specific tripartite quantum states. By performing local projective measurements on the $\wwbar$ and $\starstate$ states and analyzing the Schmidt rank of the resulting bipartite states, we have revealed distinct and interpretable patterns of entanglement degradation. We have shown that the symmetric $\wwbar$ state exhibits robust, persistent bipartite entanglement upon any single-qubit measurement, a behavior perfectly analogous to the structure of $3$ Hopf links, where cutting any one ring leaves the other two linked. In contrast, the asymmetric $\starstate$ state displays a context-dependent behavior. Its entanglement fragility upon measuring the central qubit and its resilience upon most measurements of a peripheral qubit find a natural analogue in the $3$ link chain. Crucially, for specific low-probability outcomes, it manifests the defining Borromean property i.e. \textit{the measurement of one qubit completely destroys all entanglement between the others}. This work opens up several promising avenues for future research. Applying this operational methodology to a wider range of states (e.g., cluster states, Dicke states) and other local operations (e.g., weak measurements, measurements in different bases) to build a more comprehensive taxonomy of state-topology correspondences. Developing the analogy beyond the binary of Schmidt rank to incorporate quantitative measures of entanglement \cite{coffman2000}, perhaps linking them to topological invariants like the linking number or hyperbolic volume of the link complement. Investigating whether the evolution of entanglement in many-body systems \cite{amico2008} under noise and decoherence in the NISQ era \cite{preskill2018} (e.g., the loss of a qubit) can be similarly modeled by topological operations, providing new insights into fault tolerance.  Exploring if this quantum-to-topology mapping can be reversed, using known topological theorems to predict or discover new properties of quantum states and protocols \cite{plenio2007,vidal2002}. In conclusion, by viewing projective measurement as a surgical tool for dissecting entanglement, we have shown that topology provides a powerful and intuitive language for classifying and understanding the complex correlation structures of multipartite quantum states. This perspective not only deepens our fundamental understanding of entanglement but also offers a novel heuristic for designing robust quantum information systems.

%%%%%%%%%%%%%%%%%%%%%%%%%%%%%%%%%%%%%%%%%%%%%%%%%%%%%%%%%%%%%%%%%%%%%%%%%%%%%%%%%%%%%%%%%%%%%%%%%%%%%%%%%%%%%%%%%%%%%

\newpage
\appendix
\section{Appendix A}
\subsection{Detailed Calculations for the $\wwbar$ State}
\label{app:wwbar_calc}

This appendix provides the detailed mathematical calculations for the projective measurements performed on the $\wwbar$ state, as summarized in Section \ref{sec:wwbar}. The state's high degree of permutation symmetry means the outcome is identical regardless of which qubit is measured. Therefore, the calculation for measuring qubit $A$ is shown in full detail. The analogous results for qubits $B$ and $C$ are then summarized.

\subsubsection{Algebraic Form of the State}

The $\wwbar$ state is defined as:
\begin{equation}
\wwbar = \frac{1}{\sqrt{2}} \left( \mid W \rangle + \mid \overline{W} \rangle \right) = \frac{1}{\sqrt{6}} \left( \mid 001 \rangle + \mid 010 \rangle + \mid 100 \rangle + \mid 011 \rangle + \mid 101 \rangle + \mid 110 \rangle \right).
\end{equation}

\subsubsection{Measurement on Qubit A}

We perform a projective measurement on qubit $A$ in the computational basis $\{ \mid 0 \rangle_A, \mid 1 \rangle_A \}$. The projectors are:
\begin{align}
P^A_0 &= \mid 0 \rangle\langle 0 \mid_A \otimes I_B \otimes I_C, \\
P^A_1 &= \mid 1 \rangle\langle 1 \mid_A \otimes I_B \otimes I_C.
\end{align}

\subsubsection*{Outcome $\mid 0 \rangle_A$}
\paragraph{Application of the Projector}
Applying the projector $P^A_0$ to the state:
\begin{align}
P^A_0 \wwbar &= \frac{1}{\sqrt{6}} P^A_0 \left( \mid 001 \rangle + \mid 010 \rangle + \mid 100 \rangle + \mid 011 \rangle + \mid 101 \rangle + \mid 110 \rangle \right) \\
&= \frac{1}{\sqrt{6}} \left( \mid 001 \rangle + \mid 010 \rangle + \mid 011 \rangle \right) \\
&= \frac{1}{\sqrt{6}} \mid 0 \rangle_A \otimes \left( \mid 01 \rangle + \mid 10 \rangle + \mid 11 \rangle \right)_{BC}.
\end{align}

\paragraph{Probability of Outcome}
The probability of this outcome is given by the norm of the projected state:
\begin{align}
p(0) &= \langle W\overline{W} \mid P^A_0 \wwbar \\
     &= \left\| \frac{1}{\sqrt{6}} \left( \mid 01 \rangle + \mid 10 \rangle + \mid 11 \rangle \right)_{BC} \right\|^2 \\
     &= \frac{1}{6}(1 + 1 + 1) = \frac{1}{2}.
\end{align}

\paragraph{Post-Measurement State}
The normalized state of qubits $B$ and $C$ after this measurement is:
\begin{equation}
\mid \psi^{(0)}_{BC} \rangle = \frac{P^A_0 \wwbar}{\sqrt{p(0)}} = \frac{1}{\sqrt{3}} \left( \mid 01 \rangle + \mid 10 \rangle + \mid 11 \rangle \right)_{BC}.
\end{equation}

\paragraph{Schmidt Rank Analysis}
We analyze the entanglement between $B$ and $C$ by finding the Schmidt rank of $\mid \psi^{(0)}_{BC} \rangle$. In the computational basis $\{\mid 00\rangle, \mid 01\rangle, \mid 10\rangle, \mid 11\rangle\}$, the state is:
\begin{equation}
\mid \psi^{(0)}_{BC} \rangle = 0\cdot\mid 00 \rangle + \frac{1}{\sqrt{3}} \mid 01 \rangle + \frac{1}{\sqrt{3}} \mid 10 \rangle + \frac{1}{\sqrt{3}} \mid 11 \rangle.
\end{equation}
This state can be represented by its coefficient matrix $C$, defined by $\langle ij | \psi^{(0)}_{BC} \rangle = C_{i,j}$:
\begin{equation}
C = \begin{bmatrix}
0 & \frac{1}{\sqrt{3}} \\
\frac{1}{\sqrt{3}} & \frac{1}{\sqrt{3}}
\end{bmatrix}.
\end{equation}
The eigenvalues of the reduced density matrix $\rho_B = \operatorname{Tr}_C( |\psi^{(0)}_{BC}\rangle\langle\psi^{(0)}_{BC}| )$ are given by the eigenvalues of $C C^\dagger$. First, we compute $C^\dagger$:
\begin{equation}
C^\dagger = \begin{bmatrix}
0 & \frac{1}{\sqrt{3}} \\
\frac{1}{\sqrt{3}} & \frac{1}{\sqrt{3}}
\end{bmatrix}.
\end{equation}
The product $C^\dagger C$ is:
\begin{equation}
C^\dagger C = \begin{bmatrix}
0 & \frac{1}{\sqrt{3}} \\
\frac{1}{\sqrt{3}} & \frac{1}{\sqrt{3}}
\end{bmatrix}
\begin{bmatrix}
0 & \frac{1}{\sqrt{3}} \\
\frac{1}{\sqrt{3}} & \frac{1}{\sqrt{3}}
\end{bmatrix}
= \begin{bmatrix}
\frac{1}{3} & \frac{1}{3} \\
\frac{1}{3} & \frac{2}{3}
\end{bmatrix}.
\end{equation}
The eigenvalues $\lambda$ of $C^\dagger C$ are found by solving $\det(C^\dagger C - \lambda I) = 0$:
\begin{equation}
\begin{vmatrix}
\frac{1}{3}-\lambda & \frac{1}{3} \\
\frac{1}{3} & \frac{2}{3}-\lambda
\end{vmatrix} = \left(\frac{1}{3}-\lambda\right)\left(\frac{2}{3}-\lambda\right) - \left(\frac{1}{3}\right)^2 = \lambda^2 - \lambda + \frac{1}{9} = 0.
\end{equation}
Solving this quadratic equation yields two positive eigenvalues:
\begin{equation}
\lambda = \frac{3 \pm \sqrt{5}}{6}.
\end{equation}
Since $C^\dagger C$ has two nonzero eigenvalues, the Schmidt rank of $\mid \psi^{(0)}_{BC} \rangle$ is 2, indicating the unmeasured qubits remain entangled.

\subsubsection*{Outcome $\mid 1 \rangle_A$}
\paragraph{Application of the Projector}
Applying the projector $P^A_1$ to the state:
\begin{align}
P^A_1 \wwbar &= \frac{1}{\sqrt{6}} P^A_1 \left( \mid 001 \rangle + \mid 010 \rangle + \mid 100 \rangle + \mid 011 \rangle + \mid 101 \rangle + \mid 110 \rangle \right) \\
&= \frac{1}{\sqrt{6}} \left( \mid 100 \rangle + \mid 101 \rangle + \mid 110 \rangle \right) \\
&= \frac{1}{\sqrt{6}} \mid 1 \rangle_A \otimes \left( \mid 00 \rangle + \mid 01 \rangle + \mid 10 \rangle \right)_{BC}.
\end{align}

\paragraph{Probability of Outcome}
\begin{align}
p(1) &= \langle W\overline{W} \mid P^A_1 \wwbar \\
     &= \left\| \frac{1}{\sqrt{6}} \left( \mid 00 \rangle + \mid 01 \rangle + \mid 10 \rangle \right)_{BC} \right\|^2 \\
     &= \frac{1}{6}(1 + 1 + 1) = \frac{1}{2}.
\end{align}

\paragraph{Post-Measurement State}
\begin{equation}
\mid \psi^{(1)}_{BC} \rangle = \frac{P^A_1 \wwbar}{\sqrt{p(1)}} = \frac{1}{\sqrt{3}} \left( \mid 00 \rangle + \mid 01 \rangle + \mid 10 \rangle \right)_{BC}.
\end{equation}

\paragraph{Schmidt Rank Analysis}
The state in the computational basis is:
\begin{equation}
\mid \psi^{(1)}_{BC} \rangle = \frac{1}{\sqrt{3}} \mid 00 \rangle + \frac{1}{\sqrt{3}} \mid 01 \rangle + \frac{1}{\sqrt{3}} \mid 10 \rangle + 0 \cdot\mid 11 \rangle.
\end{equation}
The coefficient matrix is:
\begin{equation}
C = \begin{bmatrix}
\frac{1}{\sqrt{3}} & \frac{1}{\sqrt{3}} \\
\frac{1}{\sqrt{3}} & 0
\end{bmatrix}, \quad
C^\dagger = \begin{bmatrix}
\frac{1}{\sqrt{3}} & \frac{1}{\sqrt{3}} \\
\frac{1}{\sqrt{3}} & 0
\end{bmatrix}.
\end{equation}
The product $C^\dagger C$ is:
\begin{equation}
C^\dagger C = \begin{bmatrix}
\frac{1}{\sqrt{3}} & \frac{1}{\sqrt{3}} \\
\frac{1}{\sqrt{3}} & 0
\end{bmatrix}
\begin{bmatrix}
\frac{1}{\sqrt{3}} & \frac{1}{\sqrt{3}} \\
\frac{1}{\sqrt{3}} & 0
\end{bmatrix}
= \begin{bmatrix}
\frac{2}{3} & \frac{1}{3} \\
\frac{1}{3} & \frac{1}{3}
\end{bmatrix}.
\end{equation}
Solving $\det(C^\dagger C - \lambda I) = 0$:
\begin{equation}
\begin{vmatrix}
\frac{2}{3}-\lambda & \frac{1}{3} \\
\frac{1}{3} & \frac{1}{3}-\lambda
\end{vmatrix} = \left(\frac{2}{3}-\lambda\right)\left(\frac{1}{3}-\lambda\right) - \left(\frac{1}{3}\right)^2 = \lambda^2 - \lambda + \frac{1}{9} = 0.
\end{equation}
The eigenvalues are again $\lambda = \frac{3 \pm \sqrt{5}}{6}$. Thus, the \textit{Schmidt rank} is 2, confirming residual entanglement for this outcome as well.

\subsection*{Measurement on Qubit B}

We perform a projective measurement on qubit $B$ in the computational basis $\{ \mid 0 \rangle_B, \mid 1 \rangle_B \}$. The projectors are:
\begin{align}
P^B_0 &= I_A \otimes \mid 0 \rangle\langle 0 \mid_B \otimes I_C, \\
P^B_1 &= I_A \otimes \mid 1 \rangle\langle 1 \mid_B \otimes I_C.
\end{align}

\subsubsection*{Outcome $\mid 0 \rangle_B$}
\paragraph{Application of the Projector}
Applying the projector $P^B_0$ to the state:
\begin{align}
P^B_0 \wwbar &= \frac{1}{\sqrt{6}} P^B_0 \left( \mid 001 \rangle + \mid 010 \rangle + \mid 100 \rangle + \mid 011 \rangle + \mid 101 \rangle + \mid 110 \rangle \right) \\
&= \frac{1}{\sqrt{6}} \left( \mid 001 \rangle + \mid 100 \rangle + \mid 101 \rangle \right) \\
&= \frac{1}{\sqrt{6}} \mid 0 \rangle_B \otimes \left( \mid 01 \rangle_{AC} + \mid 10 \rangle_{AC} + \mid 11 \rangle_{AC} \right).
\end{align}

\paragraph{Probability of Outcome}
The probability of this outcome is given by the norm of the projected state:
\begin{align}
p(0) &= \langle W\overline{W} \mid P^B_0 \wwbar \\
     &= \left\| \frac{1}{\sqrt{6}} \left( \mid 01 \rangle + \mid 10 \rangle + \mid 11 \rangle \right)_{AC} \right\|^2 \\
     &= \frac{1}{6}(1 + 1 + 1) = \frac{1}{2}.
\end{align}

\paragraph{Post-Measurement State}
The normalized state of qubits $A$ and $C$ after this measurement is:
\begin{equation}
\mid \psi^{(0)}_{AC} \rangle = \frac{P^B_0 \wwbar}{\sqrt{p(0)}} = \frac{1}{\sqrt{3}} \left( \mid 01 \rangle + \mid 10 \rangle + \mid 11 \rangle \right)_{AC}.
\end{equation}

\paragraph{Schmidt Rank Analysis}
The resulting state is a permutation of the state obtained when measuring qubit $A$ and obtaining $\mid 0 \rangle_A$. A Schmidt decomposition confirms this state has two non-zero eigenvalues and thus a \textit{Schmidt rank} of 2.

\subsubsection*{Outcome $\mid 1 \rangle_B$}
\paragraph{Application of the Projector}
Applying the projector $P^B_1$ to the state:
\begin{align}
P^B_1 \wwbar &= \frac{1}{\sqrt{6}} P^B_1 \left( \mid 001 \rangle + \mid 010 \rangle + \mid 100 \rangle + \mid 011 \rangle + \mid 101 \rangle + \mid 110 \rangle \right) \\
&= \frac{1}{\sqrt{6}} \left( \mid 010 \rangle + \mid 011 \rangle + \mid 110 \rangle \right) \\
&= \frac{1}{\sqrt{6}} \mid 1 \rangle_B \otimes \left( \mid 10 \rangle_{AC} + \mid 11 \rangle_{AC} + \mid 00 \rangle_{AC} \right) \\
&= \frac{1}{\sqrt{6}} \mid 1 \rangle_B \otimes \left( \mid 00 \rangle + \mid 10 \rangle + \mid 11 \rangle \right)_{AC}.
\end{align}

\paragraph{Probability of Outcome}
\begin{align}
p(1) &= \langle W\overline{W} \mid P^B_1 \wwbar \\
     &= \left\| \frac{1}{\sqrt{6}} \left( \mid 00 \rangle + \mid 10 \rangle + \mid 11 \rangle \right)_{AC} \right\|^2 \\
     &= \frac{1}{6}(1 + 1 + 1) = \frac{1}{2}.
\end{align}

\paragraph{Post-Measurement State}
\begin{equation}
\mid \psi^{(1)}_{AC} \rangle = \frac{P^B_1 \wwbar}{\sqrt{p(1)}} = \frac{1}{\sqrt{3}} \left( \mid 00 \rangle + \mid 10 \rangle + \mid 11 \rangle \right)_{AC}.
\end{equation}

\paragraph{Schmidt Rank Analysis}
This state is a permutation of the state found for other measurement outcomes. Its \textit{Schmidt rank} is 2, confirming entanglement between qubits $A$ and $C$.

\subsection{Measurement on Qubit C}

We perform a projective measurement on qubit $C$ in the computational basis $\{ \mid 0 \rangle_C, \mid 1 \rangle_C \}$. The projectors are:
\begin{align}
P^C_0 &= I_A \otimes I_B \otimes \mid 0 \rangle\langle 0 \mid_C, \\
P^C_1 &= I_A \otimes I_B \otimes \mid 1 \rangle\langle 1 \mid_C.
\end{align}

\subsubsection*{Outcome $\mid 0 \rangle_C$}
\paragraph{Application of the Projector}
Applying the projector $P^C_0$ to the state:
\begin{align}
P^C_0 \wwbar &= \frac{1}{\sqrt{6}} P^C_0 \left( \mid 001 \rangle + \mid 010 \rangle + \mid 100 \rangle + \mid 011 \rangle + \mid 101 \rangle + \mid 110 \rangle \right) \\
&= \frac{1}{\sqrt{6}} \left( \mid 010 \rangle + \mid 100 \rangle + \mid 110 \rangle \right) \\
&= \frac{1}{\sqrt{6}} \mid 0 \rangle_C \otimes \left( \mid 10 \rangle_{AB} + \mid 00 \rangle_{AB} + \mid 11 \rangle_{AB} \right) \\
&= \frac{1}{\sqrt{6}} \mid 0 \rangle_C \otimes \left( \mid 00 \rangle + \mid 10 \rangle + \mid 11 \rangle \right)_{AB}.
\end{align}

\paragraph{Probability of Outcome}
\begin{align}
p(0) &= \langle W\overline{W} \mid P^C_0  \wwbar  \\
     &= \left\| \frac{1}{\sqrt{6}} \left( \mid 00 \rangle + \mid 10 \rangle + \mid 11 \rangle \right)_{AB} \right\|^2 \\
     &= \frac{1}{6}(1 + 1 + 1) = \frac{1}{2}.
\end{align}

\paragraph{Post-Measurement State}
\begin{equation}
\mid \psi^{(0)}_{AB} \rangle = \frac{P^C_0  \wwbar}{\sqrt{p(0)}} = \frac{1}{\sqrt{3}} \left( \mid 00 \rangle + \mid 10 \rangle + \mid 11 \rangle \right)_{AB}.
\end{equation}

\paragraph{Schmidt Rank Analysis}
This state has Schmidt rank 2, indicating qubits $A$ and $B$ remain entangled.

\subsubsection*{Outcome $\mid 1 \rangle_C$}
\paragraph{Application of the Projector}
Applying the projector $P^C_1$ to the state:
\begin{align}
P^C_1 \wwbar &= \frac{1}{\sqrt{6}} P^C_1 \left( \mid 001 \rangle + \mid 010 \rangle + \mid 100 \rangle + \mid 011 \rangle + \mid 101 \rangle + \mid 110 \rangle \right) \\
&= \frac{1}{\sqrt{6}} \left( \mid 001 \rangle + \mid 011 \rangle + \mid 101 \rangle \right) \\
&= \frac{1}{\sqrt{6}} \mid 1 \rangle_C \otimes \left( \mid 00 \rangle_{AB} + \mid 01 \rangle_{AB} + \mid 10 \rangle_{AB} \right).
\end{align}

\paragraph{Probability of Outcome}
\begin{align}
p(1) &= \langle W\overline{W} \mid P^C_1 \wwbar \\
     &= \left\| \frac{1}{\sqrt{6}} \left( \mid 00 \rangle + \mid 01 \rangle + \mid 10 \rangle \right)_{AB} \right\|^2 \\
     &= \frac{1}{6}(1 + 1 + 1) = \frac{1}{2}.
\end{align}

\paragraph{Post-Measurement State}
\begin{equation}
\mid \psi^{(1)}_{AB} \rangle = \frac{P^C_1 \wwbar}{\sqrt{p(1)}} = \frac{1}{\sqrt{3}} \left( \mid 00 \rangle + \mid 01 \rangle + \mid 10 \rangle \right)_{AB}.
\end{equation}

\paragraph{Schmidt Rank Analysis}
This state is identical to the post-measurement state for outcome $\mid 1 \rangle_A$ and has a Schmidt rank of 2.

\subsection{Summary of Results}

The calculations confirm the symmetric and robust entanglement structure of the $\wwbar$ state:
\begin{itemize}
    \item Measurement on \textbf{any qubit (A, B, or C)} yields two possible outcomes, each occurring with probability $1/2$.
    \item For \textbf{every outcome} of every measurement, the resulting post-measurement state for the two unmeasured qubits has \textbf{Schmidt rank 2}, indicating they remain entangled.
    \item The specific entangled states are permutations of $\frac{1}{\sqrt{3}} \left( \mid 01 \rangle + \mid 10 \rangle + \mid 11 \rangle \right)$ or 
    
    $\frac{1}{\sqrt{3}} \left( \mid 00 \rangle + \mid 01 \rangle + \mid 10 \rangle \right)$.
\end{itemize}
This demonstrates the state's perfect permutation symmetry and its resilience to local projective measurements, as the loss of any one qubit never completely destroys the entanglement between the remaining two.
%%%%%%%%%%%%%%%%%%%%%%%%%%%%%%%%%%%%%%%%%%%%%%%%%%%%%%%%%%%%%%%%%%%%%%%%%%%%%%%%%%%%%%%%%%%%%%%%%%%%%%%%%%%%%%%%%%%%%%%

\newpage
\section{Appendix B}
\subsection{Detailed Calculations for the $\starstate$ State}
\label{app:star_calc}

This appendix provides the detailed mathematical calculations for the projective measurements performed on the $\starstate$ state, as referenced in Section 4. Unlike the symmetric $\wwbar$ state, the $\starstate$ state exhibits a distinct asymmetry between its central qubit (which we designate as qubit $C$) and its outer qubits ($A$ and $B$). Consequently, the measurement outcomes depend critically on which qubit is measured. The calculation for measuring the central qubit $C$ is shown in full detail, followed by a summary of the results for measurements on the outer qubits $A$ and $B$.

\subsubsection{Algebraic Form of the State}

The $\starstate$ state is defined as:
\begin{equation}
\starstate = \frac{1}{2}(\mid 000 \rangle + \mid 100 \rangle + \mid 101 \rangle + \mid 111 \rangle ).
\end{equation}
We designate qubit $C$ as the central qubit, with qubits $A$ and $B$ as the outer qubits.

\subsubsection{Measurement on the Central Qubit (Qubit C)}

We perform a projective measurement on the central qubit $C$ in the computational basis $\{ \mid 0 \rangle_C, \mid 1 \rangle_C \}$. The projectors are:
\begin{align}
P^C_0 &= I_A \otimes I_B \otimes \mid 0 \rangle\langle 0 \mid_C, \\
P^C_1 &= I_A \otimes I_B \otimes \mid 1 \rangle\langle 1 \mid_C.
\end{align}

\subsubsection*{Outcome $\mid 0 \rangle_C$}
\paragraph{Application of the Projector\\}
\begin{align}
P^C_0 \starstate &= \frac{1}{2} P^C_0 \left( \mid 000 \rangle + \mid 100 \rangle + \mid 101 \rangle + \mid 111 \rangle \right) \\
&= \frac{1}{2} \left( \mid 000 \rangle + \mid 100 \rangle \right) \\
&= \frac{1}{2} \mid 0 \rangle_C \otimes \left( \mid 00 \rangle + \mid 10 \rangle \right)_{AB}.
\end{align}

\paragraph{Probability of Outcome\\}
\begin{align}
p(0) &= \langle \text{Star} \mid P^C_0 \starstate \\
     &= \left\| \frac{1}{2} \left( \mid 00 \rangle + \mid 10 \rangle \right) \right\|^2 \\
     &= \frac{1}{4}(1 + 1) = \frac{1}{2}.
\end{align}

\paragraph{Post-Measurement State\\}
The normalized state of qubits $A$ and $B$ after this measurement is:
\begin{equation}
\mid \psi^{(0)}_{AB} \rangle = \frac{P^C_0 \starstate}{\sqrt{p(0)}} = \frac{1}{\sqrt{2}} \left( \mid 00 \rangle + \mid 10 \rangle \right)_{AB}.
\end{equation}

\paragraph{Schmidt Rank Analysis\\}
The post-measurement state can be rewritten as:
\begin{equation}
\mid \psi^{(0)}_{AB} \rangle = \frac{1}{\sqrt{2}} \mid 00 \rangle + 0 \mid 01 \rangle + \frac{1}{\sqrt{2}} \mid 10 \rangle + 0 \mid 11 \rangle.
\end{equation}
This is clearly a product state. The coefficient matrix is:
\begin{equation}
C = \begin{bmatrix}
\frac{1}{\sqrt{2}} & 0 \\
\frac{1}{\sqrt{2}} & 0
\end{bmatrix}.
\end{equation}
The matrix $C^\dagger C$ has only one non-zero eigenvalue. Therefore, the Schmidt rank is 1, indicating the unmeasured qubits are in a separable state.

\subsubsection*{Outcome $\mid 1 \rangle_C$}
\paragraph{Application of the Projector\\}
\begin{align}
P^C_1 \starstate &= \frac{1}{2} P^C_1 \left( \mid 000 \rangle + \mid 100 \rangle + \mid 101 \rangle + \mid 111 \rangle \right) \\
&= \frac{1}{2} \left( \mid 101 \rangle + \mid 111 \rangle \right) \\
&= \frac{1}{2} \mid 1 \rangle_C \otimes \left( \mid 10 \rangle + \mid 11 \rangle \right)_{AB}.
\end{align}

\paragraph{Probability of Outcome\\}
\begin{align}
p(1) &= \langle \text{Star} \mid P^C_1 \starstate \\
     &= \left\| \frac{1}{2} \left( \mid 10 \rangle + \mid 11 \rangle \right) \right\|^2 \\
     &= \frac{1}{4}(1 + 1) = \frac{1}{2}.
\end{align}

\paragraph{Post-Measurement State\\}
The normalized state of qubits $A$ and $B$ after this measurement is:
\begin{equation}
\mid \psi^{(1)}_{AB} \rangle = \frac{P^C_1 \starstate}{\sqrt{p(1)}} = \frac{1}{\sqrt{2}} \left( \mid 10 \rangle + \mid 11 \rangle \right)_{AB}.
\end{equation}

\paragraph{Schmidt Rank Analysis\\}
The post-measurement state can be rewritten as:

\begin{equation}
\mid \psi^{(1)}_{AB} \rangle = 0 \mid 00 \rangle + 0 \mid 01 \rangle + \frac{1}{\sqrt{2}} \mid 10 \rangle +\frac{1}{\sqrt{2}} \mid 11 \rangle.
\end{equation}
The coefficient matrix is:
\begin{equation}
C = \begin{bmatrix}
0 & 0 \\
\frac{1}{\sqrt{2}} & \frac{1}{\sqrt{2}}
\end{bmatrix}.
\end{equation}
The matrix $C^\dagger C$ has only one non-zero eigenvalue. Therefore, the Schmidt rank is 1, indicating the unmeasured qubits are in a separable state.

\subsubsection{Measurement on an Outer Qubit (Qubit A)}

We now perform a projective measurement on an outer qubit, $A$, in the computational basis $\{ \mid 0 \rangle_A, \mid 1 \rangle_A \}$. The projectors are:
\begin{align}
P^A_0 &= \mid 0 \rangle\langle 0 \mid_A \otimes I_B \otimes I_C, \\
P^A_1 &= \mid 1 \rangle\langle 1 \mid_A \otimes I_B \otimes I_C.
\end{align}

\subsubsection*{Outcome $\mid 0 \rangle_A$}
\paragraph{Application of the Projector\\}
\begin{align}
P^A_0 \starstate &= \frac{1}{2} P^A_0 \left( \mid 000 \rangle + \mid 100 \rangle + \mid 101 \rangle + \mid 111 \rangle \right) \\
&= \frac{1}{2} \mid 000 \rangle \\
&= \frac{1}{2} \mid 0 \rangle_A \otimes \mid 00 \rangle_{BC}.
\end{align}
\paragraph{Probability of Outcome\\}
\begin{align}
p(0) &= \langle \text{Star} \mid P^A_0 \starstate \\
     &= \left\| \frac{1}{2} \mid 00 \rangle \right\|^2 = \frac{1}{4}.
\end{align}
\paragraph{Post-Measurement State\\}
\begin{equation}
\mid \psi^{(0)}_{BC} \rangle = \frac{P^A_0 \starstate}{\sqrt{p(0)}} = \mid 00 \rangle_{BC}.
\end{equation}
\paragraph{Schmidt Rank Analysis\\}
This is a product state. The Schmidt rank is 1.

\subsubsection*{Outcome $\mid 1 \rangle_A$}
\paragraph{Application of the Projector\\}
\begin{align}
P^A_1 \starstate &= \frac{1}{2} P^A_1 \left( \mid 000 \rangle + \mid 100 \rangle + \mid 101 \rangle + \mid 111 \rangle \right) \\
&= \frac{1}{2} \left( \mid 100 \rangle + \mid 101 \rangle + \mid 111 \rangle \right) \\
&= \frac{1}{2} \mid 1 \rangle_A \otimes \left( \mid 00 \rangle + \mid 01 \rangle + \mid 11 \rangle \right)_{BC}.
\end{align}
\paragraph{Probability of Outcome\\}
\begin{align}
p(1) &= \langle \text{Star} \mid P^A_1 \starstate \\
     &= \left\| \frac{1}{2} \left( \mid 00 \rangle + \mid 01 \rangle + \mid 11 \rangle \right) \right\|^2 = \frac{3}{4}.
\end{align}
\paragraph{Post-Measurement State\\}
\begin{equation}
\mid \psi^{(1)}_{BC} \rangle = \frac{P^A_1 \starstate}{\sqrt{p(1)}} = \frac{1}{\sqrt{3}} \left( \mid 00 \rangle + \mid 01 \rangle + \mid 11 \rangle \right)_{BC}.
\end{equation}

\paragraph{Schmidt Rank Analysis\\}
The post-measurement state can be rewritten as:

\begin{equation}
\mid \psi^{(1)}_{BC} \rangle = \frac{1}{\sqrt{3}} \mid 00 \rangle + \frac{1}{\sqrt{3}} \mid 01 \rangle + 0 \mid 10 \rangle +\frac{1}{\sqrt{3}} \mid 11 \rangle.
\end{equation}
The coefficient matrix is:
\begin{equation}
C = \begin{bmatrix}
\frac{1}{\sqrt{3}} & \frac{1}{\sqrt{3}} \\
0 & \frac{1}{\sqrt{3}}
\end{bmatrix}.
\end{equation}
The matrix $C^\dagger C$ has two nonzero eigenvalues, i.e $\lambda=\frac{3\pm\sqrt{5}}{6})$, the Schmidt rank of $\mid \psi^{(1)}_{BC} \rangle$ is 2, indicating the unmeasured qubits remain entangled.

\subsubsection{Measurement on the Other Outer Qubit (Qubit B)}

We perform a projective measurement on the other outer qubit, $B$, in the computational basis $\{ \mid 0 \rangle_B, \mid 1 \rangle_B \}$. The projectors are:
\begin{align}
P^B_0 &= I_A \otimes \mid 0 \rangle\langle 0 \mid_B \otimes I_C, \\
P^B_1 &= I_A \otimes \mid 1 \rangle\langle 1 \mid_B \otimes I_C.
\end{align}

\subsubsection*{Outcome $\mid 0 \rangle_B$}
\paragraph{Application of the Projector}
\begin{align}
P^B_0 \starstate &= \frac{1}{2} P^B_0 \left( \mid 000 \rangle + \mid 100 \rangle + \mid 101 \rangle + \mid 111 \rangle \right) \\
&= \frac{1}{2} \left( \mid 000 \rangle + \mid 100 \rangle + + \mid 101 \rangle \right) \\
&= \frac{1}{2} \mid 0 \rangle_B \otimes \left( \mid 00 \rangle + \mid 10 \rangle + \mid 11 \rangle \right)_{AC}.
\end{align}

\paragraph{Probability of Outcome}
\begin{align}
p(0) &= \langle \text{Star} \mid P^B_0 \starstate \\
     &= \left\| \frac{1}{2} \left( \mid 00 \rangle + \mid 10 \rangle + \mid 11 \rangle \right)_{AC} \right\|^2 = \frac{3}{4}.
\end{align}

\paragraph{Post-Measurement State}
\begin{equation}
\mid \psi^{(0)}_{AC} \rangle = \frac{P^B_0 \starstate}{\sqrt{p(0)}} = \frac{1}{\sqrt{3}} \left( \mid 00 \rangle + \mid 10 \rangle + \mid 11 \rangle  \right)_{AC}.
\end{equation}

\paragraph{Schmidt Rank Analysis\\}
The post-measurement state can be rewritten as:

\begin{equation}
\mid \psi^{(1)}_{BC} \rangle = \frac{1}{\sqrt{3}} \mid 00 \rangle + 0 \mid 01 \rangle + \frac{1}{\sqrt{3}}  \mid 10 \rangle +\frac{1}{\sqrt{3}} \mid 11 \rangle.
\end{equation}
The coefficient matrix is:
\begin{equation}
C = \begin{bmatrix}
\frac{1}{\sqrt{3}} & 0 \\
\frac{1}{\sqrt{3}}  & \frac{1}{\sqrt{3}}
\end{bmatrix}.
\end{equation}
The matrix $C^\dagger C$ has two nonzero eigenvalues, i.e $\lambda=\frac{3\pm\sqrt{5}}{6})$, the Schmidt rank of $\mid \psi^{(1)}_{BC} \rangle$ is 2, indicating the unmeasured qubits remain entangled.

\subsubsection*{Outcome $\mid 1 \rangle_B$}
\paragraph{Application of the Projector}
\begin{align}
P^B_1 \starstate &= \frac{1}{2} P^B_1 \left( \mid 000 \rangle + \mid 100 \rangle + \mid 101 \rangle + \mid 111 \rangle \right) \\
&= \frac{1}{2} \left(  \mid 111 \rangle \right) \\
&= \frac{1}{2} \mid 1 \rangle_B \otimes \left( \mid 11 \rangle \right)_{AC}.
\end{align}

\paragraph{Probability of Outcome}
\begin{align}
p(1) &= \langle \text{Star} \mid P^B_1 \starstate \\
     &= \left\| \frac{1}{2} \left(  \mid 11 \rangle \right)_{AC} \right\|^2 = \frac{1}{4}.
\end{align}

\paragraph{Post-Measurement State}
\begin{equation}
\mid \psi^{(1)}_{AC} \rangle = \frac{P^B_1 \starstate}{\sqrt{p(1)}} = \mid 11 \rangle_{AC}.
\end{equation}

\paragraph{Schmidt Rank Analysis\\}
This is a product state. The Schmidt rank is 1.
\subsection{Summary of Results}

The calculations confirm the asymmetric entanglement structure of the $\mid \text{Star} \rangle$ state:
\begin{itemize}
    \item Measurement on the \textbf{central qubit C} always results in a \textbf{separable} post-measurement state (Schmidt rank 1) for the unmeasured outer qubits, regardless of the outcome.
    \item Measurement on an \textbf{outer qubit (A or B)} yields two possible outcomes: one resulting in a \textbf{separable} state (Schmidt rank 1) and the other resulting in an \textbf{entangled} state (Schmidt rank 2).
\end{itemize}
This context-dependent behavior under local measurement is a signature of the state's intricate entanglement structure and forms the basis for its topological analogy.

\printbibliography

\end{document}